\documentclass[aip,jcp,showpacs,twocolumn,superscriptaddress,amsmath,amssymb,reprint,floatfix]{revtex4-1}
\usepackage{amssymb}
\usepackage{amsmath}
\usepackage{graphicx}% Include figure files
\usepackage{dcolumn}% Align table columns on decimal point
\usepackage{bm}% bold math
\usepackage{url}
\usepackage{setspace}
\usepackage{multirow}
\usepackage[pdftex,colorlinks=true,	linkcolor=red,	citecolor=blue,	urlcolor=blue]{hyperref}

\begin{document}

\preprint{AIP/123-QED}

\title{Action spectroscopy of SrCl$^+$ using an integrated ion trap time-of-flight mass spectrometer}

\author{Prateek Puri}
\email{teek24@ucla.edu}
\author{Steven J. Schowalter}
\affiliation{Department of Physics and Astronomy, University of California, Los Angeles, California 90095, USA}
\author{Svetlana Kotochigova}
\affiliation{Department of Physics, Temple University, Philadelphia, Pennsylvania 19122, USA}
\author{Alexander Petrov}
\altaffiliation{Alternative address: St. Petersburg Nuclear Physics Institute, Gatchina, 188300; Division of Quantum Mechanics, St. Petersburg State University, 198904, Russia.}
\affiliation{Department of Physics, Temple University, Philadelphia, Pennsylvania 19122, USA}
\author{Eric R. Hudson}
\affiliation{Department of Physics and Astronomy, University of California, Los Angeles, California 90095, USA}
\date{\today}

\begin{abstract}
The photodissociation cross-section of SrCl$^+$ is measured in the spectral range  of 36000 -- 46000~cm$^{-1}$ using a modular time-of-flight mass spectrometer (TOF-MS). By irradiating a sample of trapped SrCl$^+$ molecular ions with a pulsed dye laser, X$^1\Sigma^+$ state molecular ions are electronically excited to the repulsive wall of the A$^1\Pi$ state, resulting in dissociation. Using the TOF-MS, the fragments are detected and the photodissociation cross-section is determined for a broad range of photon energies. Detailed \textit{ab initio} calculations of the molecular potentials and spectroscopic constants are also performed and are found to be in good agreement with experiment. The spectroscopic constants for SrCl$^+$ are also compared to those of another alkaline earth chalcogen, BaCl$^+$, in order to highlight structural differences between the two molecular ions. This work represents the first spectroscopy and \textit{ab initio} calculations of SrCl$^+$.
\end{abstract}

\pacs{}% PACS, the Physics and Astronomy
                             % Classification Scheme.
\keywords{photodissociation, linear quadrupole trap, molecular ion}%Use showkeys class option if keyword
                              %display desired
\maketitle

\section{Introduction}

Diatomic molecular ions hold immense promise for advances in both the fundamental and applied sciences. The internal structure of these species is complex relative to that of atoms, giving rise to rich physics and chemistry, yet their structure remains simple enough to allow control over all degrees of freedom~\cite{Staanum2010a,Schneider2010,Tong2010,Rellergert2013a}. Already, work is underway using these ions to study chemical reactions at the quantum level ~\cite{Rellergert2011a,Sullivan2011,Sullivan2012,Hall2012,Tong2012}, to understand important astrophysical processes \cite{McEwan1999,Snow2008,Reddy2010,Indriolo2010}, to perform precision measurement tests of fundamental physics~\cite{Meyer2006,Cossel2012,Bressel2012,Loh2013,Bakalov2013}, and to implement quantum logic operations~\cite{Mur-Petit2012a,Goeders2013,Shi2013}. Molecular ions may also play a central role in the development of scalable quantum computation architectures~\cite{DeMille2002,Andre2006,Schuster2011}. Despite these important efforts to study and control diatomic molecular ions, progress has been slowed by one simple fact: \textit{there is very little spectroscopic data available for molecular ions}.

The paucity of molecular ion spectroscopy is most likely attributed to several factors which conspire to make this spectroscopy more difficult than that of neutral molecules. First, molecular ions often exhibit short experimental lifetimes due to fast ion-molecule reactions and rapid diffusion under the influence of small electric fields~\cite{Saykally1981}. Second, the Coulomb repulsion between ions typically limits available densities\cite{Douglas2005} to $\sim10^8$~cm$^{-3}$. Further, though a systematic review of the available spectroscopic data for small molecular ions was carried out by Berkowitz
and Groeneveld~\cite{Berkowitz1983} in 1983 in an effort to encourage the community, in recent years the majority of interest -- with notable exceptions \cite{Heaven2011,Kreckel2012,Siller2013,Stewart2013,Merritt2009,Han2011,Hodges2013} -- has shifted toward large molecular ions~\cite{Casaburi2009}, atomic and molecular clusters~\cite{Last2004}, and multiply charged ions~\cite{Duncan2000}. Therefore, for the burgeoning field of ultracold molecular-ion research to realize its full potential, new efforts in small molecular ion spectroscopy are required.\\
\indent Here we present the first spectroscopic data to date of the molecular ion SrCl$^+$. By trapping the molecular ions in a linear quadropole trap (LQT), the experimental lifetime of the molecular ions is extended, allowing them to be interrogated on convenient timescales. The A$^1\Pi\leftarrow$X$^1\Sigma^+$ dissociation cross-section is measured using a time-of-flight mass spectrometer (TOF-MS), and the corresponding molecular potentials and spectroscopic constants are calculated. By employing action spectroscopy, the sensitive techniques of mass spectrometry and ion detection can be used to mitigate the effect of small ion sample sizes. The motivation and composition of this report are similar to that of the first spectroscopic study~\cite{Chen2011} of another alkaline earth chalcogen, BaCl$^+$. Due to the similarity of their rovibronic structures, SrCl$^+$ is a candidate for ultracold molecular ion experiments similar to those~\cite{Rellergert2013a} currently underway using BaCl$^+$. In addition, the experimentally convenient ground state rotational splitting of $\sim6$~GHz makes SrCl$^+$ a potentially attractive qubit for quantum information studies~\cite{Schuster2011}.\\
\indent In the remainder of this manuscript, we outline the experimental apparatus, explain the spectroscopy protocol, and present the first spectroscopic data for SrCl$^+$ photodissociation. We also present the results of an \textit{ab initio} calculation of the SrCl$^+$ structure, which show good agreement with the experimental data. We conclude with a comparison of SrCl$^+$ structure to that of BaCl$^+$, which we have also recently determined\cite{Rellergert2013a}.

\section{Experimental Design}

\begin{figure}[b]
\resizebox{90mm}{!}{
    \includegraphics{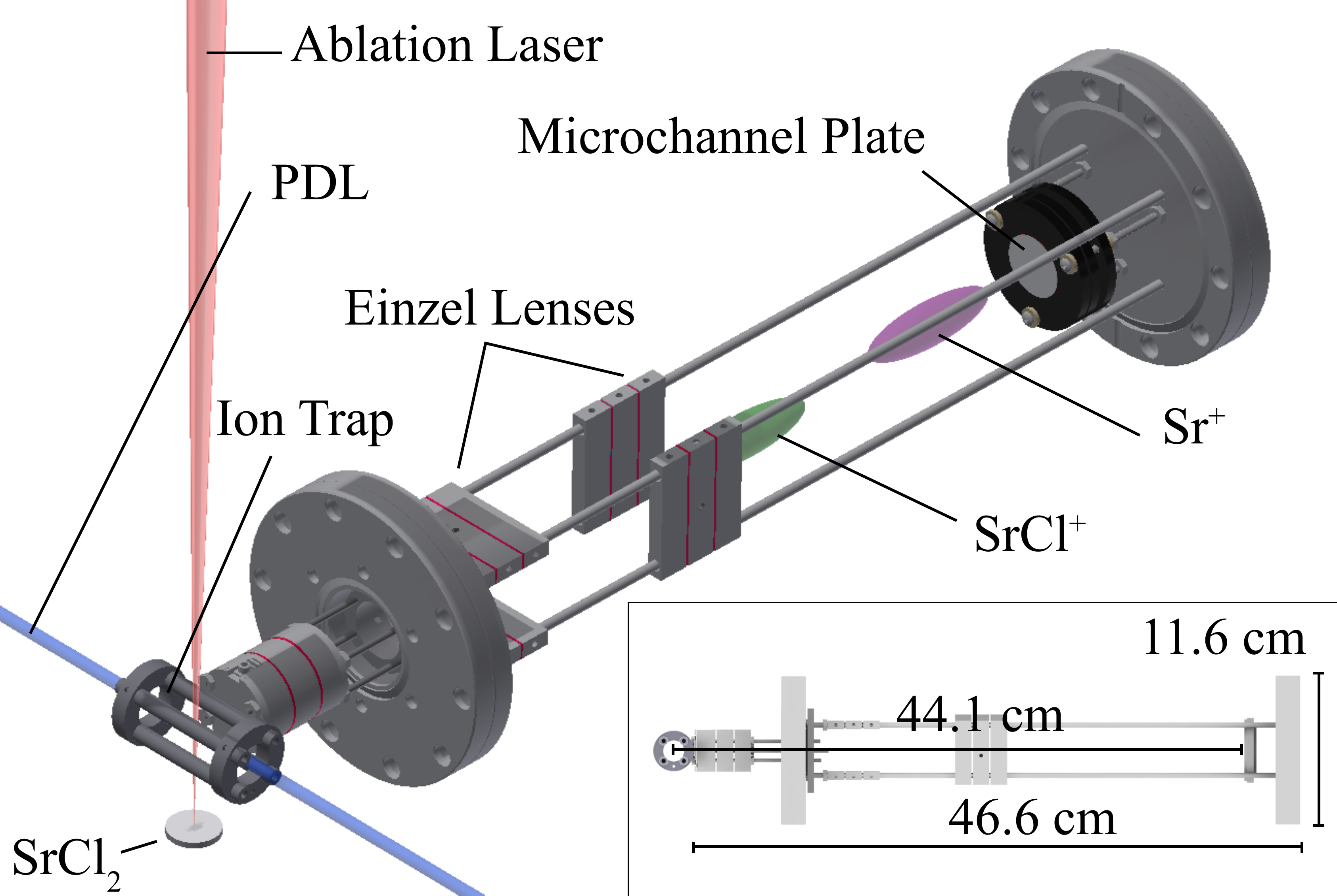}
}  \caption{A 3D rendering of the integrated ion trap and TOF-MS with insets showing relevant lengths. (Color online) \label{fig:tof}}
\end{figure}

The apparatus used to perform the action spectroscopy of SrCl$^+$, shown in Figure~\ref{fig:tof}, consists of a LQT coupled to a modular TOF-MS similar to that reported in Ref. \onlinecite{Schowalter2012}.  This device is housed in a vacuum chamber maintained at a background pressure of 10$^{-8}$~mbar. A pressed, annealed SrCl$_2$ pellet is mounted below the trap. The LQT has a field radius, $r_0$, of 11.1~mm and an electrode radius, $r_e$, of 6.35~mm.  A trapping radiofrequency (rf) voltage is applied to all four electrodes with typical amplitude $V_{\mathrm{rf}}=120$~V and frequency $\Omega=2\pi\times400$~kHz, resulting in a Mathieu $q$-parameter of $q= (4QV_{\mathrm{rf}})/(mr_0^2\Omega^2)\approx0.48$, where $Q$ is the charge and $m$ is the mass of SrCl$^+$. The mass spectrum of trapped ions is recorded by using a pulsed, high-voltage scheme described in Ref. \onlinecite{Schowalter2012}, which creates a two-stage electric field that ejects the ions \cite{Wiley1955} into a radially-oriented TOF-MS.

The  TOF-MS used in this experiment is a modified version of a device used previously~\cite{Chen2011,Schowalter2012,Rellergert2013a} for spectroscopic studies of BaCl$^+$. The primary modifications to this TOF-MS include the replacement of a channel electron multiplier with a micro-channel plate, the transition from spherical to cylindrical Einzel lenses, the lengthening of the field-free drift tube from $25.2$~cm to $44.1$~cm, and the inclusion of an adjustable skimmer. With these modifications, the mass resolution of the TOF-MS is experimentally verified to be a factor of $\sim$ $2$ larger, and simulation has shown the detection efficiency is improved by an order of magnitude.

\begin{figure}[t]
\resizebox{90mm}{!}{
    \includegraphics{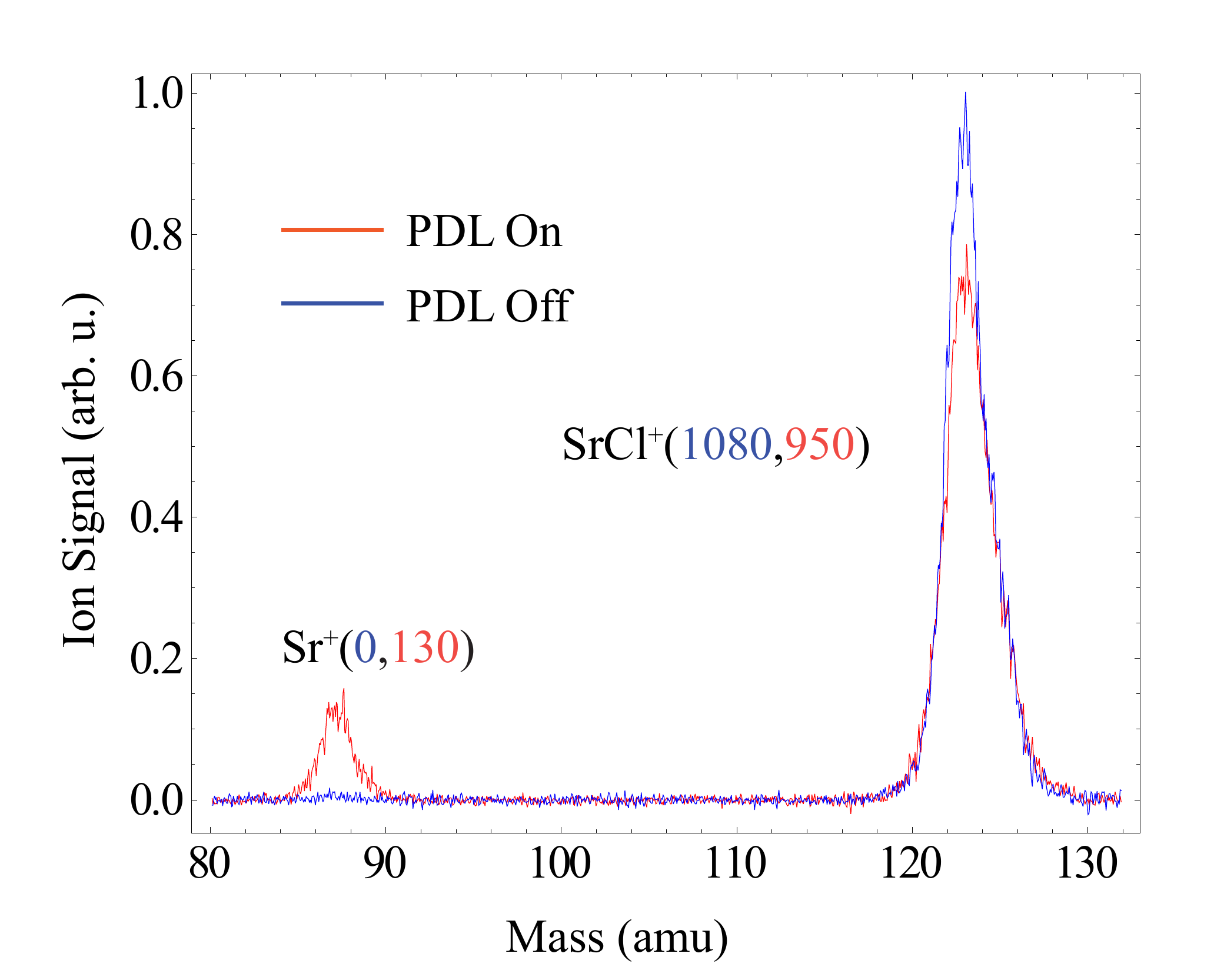}
}  \caption{The TOF mass spectra for trapped SrCl$^+$ with (red) and without (blue) the presence of the PDL tuned to 44006.78~cm$^{-1}$. In parenthesis, ion number is presented for each species when the PDL is off and on, respectively. (Color online) \label{fig:trace}}
\end{figure}

\section{Action Spectroscopy}

The SrCl$^+$ photodissociation cross-section is measured as follows. The SrCl$_2$ pellet is ablated with a pulsed laser (Nd:YAG, 1064 nm, $\sim$1 mJ, $\sim$10 ns pulse), which creates a plume of fast-moving charged and neutral species. The LQT parameters are set such that only SrCl$^+$ or ions with a larger mass-to-charge ratio can be loaded\cite{Hashimoto2006} into the trap; however, SrCl$^+$ is the only species detected following this loading process. The V$_{\mathrm{rf}}$ is ramped from $120$~V to $60$~V in $150$~ms, a value where both SrCl$^+$ and any fragment Sr$^+$ produced by photodissociation can be co-trapped. While a room-temperature helium buffer gas has been used to collisionally cool trapped ions and improve the TOF-MS detection efficiency, with this improved apparatus the detection efficiency is adequate without buffer gas and therefore not used.

Once the trapped sample of SrCl$^+$ is initialized, the ions are exposed to pulses of light propagating along the axis of the LQT with a repetition rate $r=10$~Hz for $1.1$~s. This light is the frequency doubled output of a pulsed dye laser (PDL) pumped by $355$~nm light from a tripled, pulsed Nd:YAG laser.  The doubled PDL light can be tuned over the entire A$^1\Pi\leftarrow$X$^1\Sigma^+$ photodissociation cross-section ($36000$~-~$46000$~cm$^{-1}$) using four different Coumarin dyes. To mitigate sharp intensity fluctuations due to pointing instability, the beam is expanded to approach a roughly uniform intensity profile ($\sim$ $5$~mm in diameter) over the spatial extent of the ion cloud. Beam overlap with the ion cloud is achieved such that the photodissociation rate is maximized for each measurement. The pulse energy, $E$, is measured by an energy meter located near the exit viewport of the vacuum chamber and varies between 0.5~mJ and 2~mJ over the gain profiles of the laser dyes used, with a typical value of $\sim1$~mJ.

Following the photodissociation sequence, the trapping rf is switched off in $\sim 5$~$\mu$s and high-voltage extraction pulses ($2.0$ and $1.8$~kV) are applied with a $<$ $1$~$\mu$s rise time to laterally-paired LQT rods, which eject the ions into the TOF-MS~\cite{Schowalter2012}. Figure~\ref{fig:trace} shows typical mass spectra with and without exposure to the PDL light, with observable Sr$^+$ production in the former case. Mass spectra without the PDL are recorded prior to each measurement as a control to ensure that detected Sr$^+$ ions are produced solely by photodissociation. Furthermore, due to an occasional reaction between trapped Sr$^+$ fragments and background gas, SrOH$^+$ peaks are also included in the photodissociation fraction, similar to a reaction involving Ba$^+$ previously observed in Ref. \onlinecite{Schowalter2012}.   

\begin{figure}[t]
\resizebox{90mm}{!}{
    \includegraphics{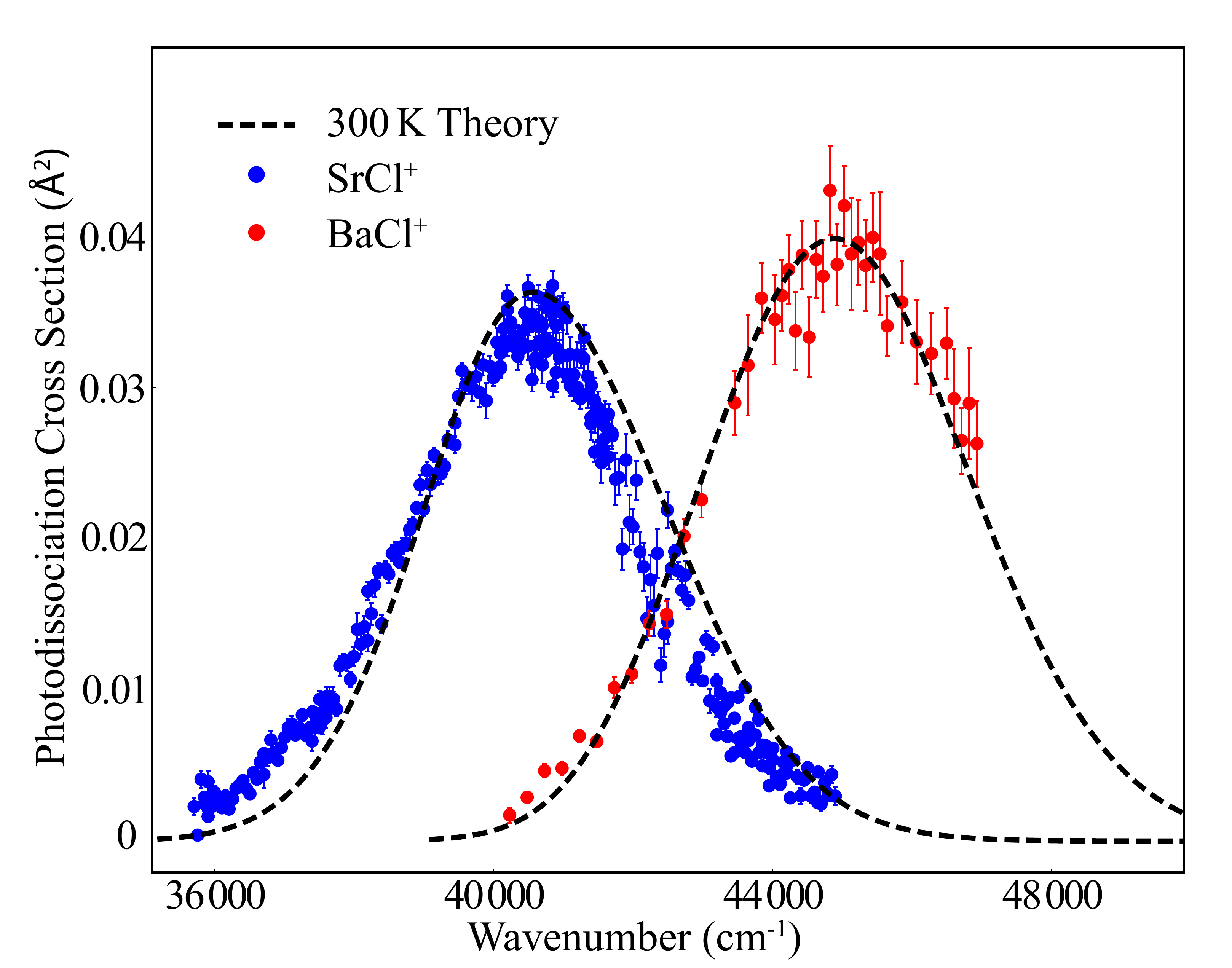}
}  \caption{Cross-section of the A$^1\Pi\leftarrow$X$^1\Sigma^+$ photodissociation transition in SrCl$^+$ at $300$~K. The $300$~K theoretical cross-section is also presented as a dashed line. A peak value of $0.0367$~$\mathrm{\AA}^2$ is measured at $40481$~cm$^{-1}$ For comparison, both the experimental and the theoretical $300$~K cross-sections for the A$^1\Pi\leftarrow$X$^1\Sigma^+$ photodissociation transition in BaCl$^+$ are also displayed\cite{Rellergert2013a}. (Color online) \label{fig:cs}}
\end{figure}

By measuring the amount of Sr$^+$ relative to the initial amount of SrCl$^+$, a photodissociation fraction, $\eta$ can be extracted and used to calculate the photodissociation cross-section, $\sigma$, at wavenumber $K$, as:

\begin{align}
\sigma = \frac{h c K}{\bar{I} t}\ln{\frac{1}{(1-\eta)}}
\end{align}
where $h$ is Planck's constant, $c$ is the speed of light, and $\overline{I}$ is the effective PDL intensity given by $(E \times r)/A$, where $A$ is the beam area. By repeating the process at each wavenumber $K$ the photodissociation cross-section is measured in steps of $50$ cm$^{-1}$ as shown in Figure~\ref{fig:cs}. Each point in this data set represents the average of ten experimental cycles, each $\sim$3~s in duration. Statistical error in the experiment, identified by the error bars, arises in the measurement of beam intensity, pulse energy, and photodissociation fraction at each PDL wavenumber. Typical statistical error is determined to be $\lesssim 6$\%  for each data point. Long timescale drifts in the PDL beam shape and intensity makes it necessary to take data in intervals of $\sim$$1000$~cm$^{-1}$. To calibrate each interval, cross-section values are remeasured and normalized to those of previously measured intervals. This normalization factor is typically found to be between $0.3$ and $1$. To calibrate the absolute magnitude of the cross-section, we normalize this data to the known photodissociation cross-section of BaCl$^+$ \cite{Rellergert2013a}(shown in Figure~\ref{fig:cs}), which we also measure with this apparatus. This normalization factor is found to be $2.65$, and along with the interval normalization, is the dominant source of systematic error, which is estimated to be $\leq 10 \times$.

\section{Theory}

\begin{figure}
\includegraphics[scale=0.45]{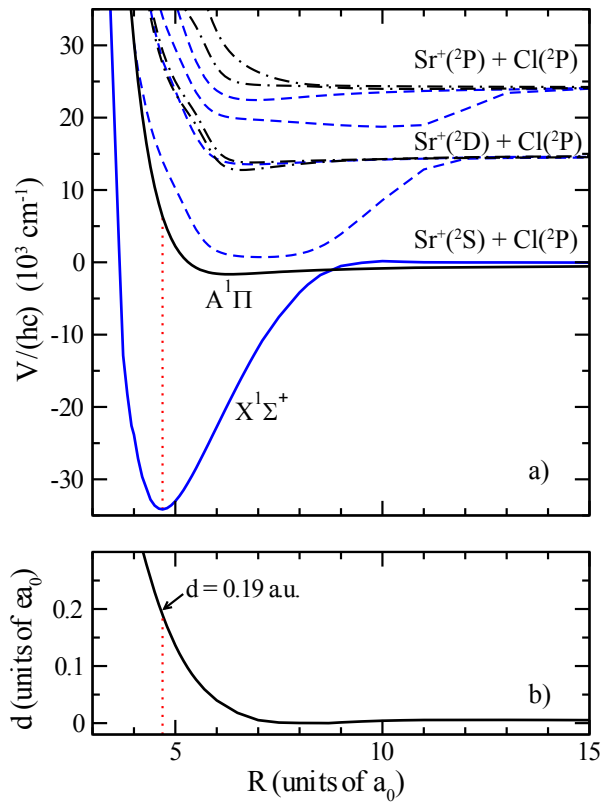}
\caption{Panel a): Molecular potentials of the SrCl$^+$ molecule as
a function of internuclear separation $R$. Solid curves indicate potentials that are involved in the photodissociation and are labeled by their X$^1\Sigma^+$ and A$^1\Pi$ symmetry. Other potentials are shown by dashed and dashed--dotted lines for the $^1\Sigma^+$ (black) and $^1\Pi$ (blue) symmetries, respectively. The vertical line indicates a transition from the lowest rovibrational
level of the ground-state potential to a continuum state of the A$^1\Pi$
potential driven by one-color laser radiation. Panel b): The  X$^1\Sigma^+$ to A$^1\Pi$ electronic transition dipole 
moment as a function of internuclear separation $R$. The dipole moment at the equilibrium 
separation of the X$^1\Sigma^+$  state is indicated. (Color online)}
\label{curves}
\end{figure}

To identify this photodissociation pathway, we have calculated the $^1\Sigma$ and $^1\Pi$ non-relativistic electronic potentials of the SrCl$^+$ molecule dissociating to the Sr$^+$($^2S$)+Cl($^2P$), Sr$^+$($^2D$)+Cl($^2P$), and Sr$^+$($^2P$)+Cl($^2P$) limits, shown in Figure~\ref{curves}a. The potential energies, transition and permanent dipole moments for the X$^1\Sigma^+$ and A$^1\Pi$ potentials, which are relevant for our experimental measurements, have been calculated using the coupled cluster method with single, double, and perturbative triple excitations (CCSD(T)) for the ground X state and coupled cluster equation of motion (EOM-CCSD) for the excited A state. The def2-QZVPP basis sets for Cl:(20s14p4d2f1g)/[9s6p4d2f1g] and Sr:(8s8p5d3f)/[7s5p4d3f]~\cite{Weigend2005} with the Stuttgart ECP28MDF relativistic effective core potential~\cite{Lim2005} are used.

For completeness, higher excited states have been calculated based on the multi-reference configuration interaction (MRCI) method in the MOLPRO software suite. The Gaussian basis set aug-cc-pVTZ [6s,5p,3d,2f]~\cite{Woon1993} for Cl and the correlation consistent [8s8p5d4f] with ECP28MDF pseudo-potential~\cite{Lim2006} for Sr are applied. Reference configurations are obtained in a complete active space self-consistent field calculation with $5s,5p,5d$ orbitals for Sr and $3p,3d$ orbitals for Cl. The core electrons $4s^24p^6$ for Sr and $1s^22s^22p^6$ for Cl are not correlated in the MRCI calculation. The potential curves that are relevant to the experimental observation have solid lines and are labeled as X$^1\Sigma^+$and A$^1\Pi$ for the ground and first excited state, respectively.

Using the CCSD(T) and EOM-CSSD methods described above, the electronic dipole matrix element as a function of internuclear separation $R$ for the A$^1\Pi\leftarrow$ X$^1\Sigma^+$ transition is also calculated and shown in Figure~\ref{curves}b. With this dipole moment function and the molecular potentials for these states, LeRoy$ ' $s BCONT program~\cite{Roy2004} is used to calculate a thermally-averaged theoretical photodissociation cross-section shown in Fig.~\ref{fig:cs}. For this calculation, it is assumed that the SrCl$^+$ rovibrational degrees of freedom are in equilibrium with the $300$~K blackbody radiation of the vacuum chamber. As seen in Fig.~\ref{fig:cs}, the agreement between the measured and predicted cross-section is good. Given that we estimate a systematic error in the absolute cross-section measurement of $\leq 10$$\times$, the similarity of the measured and calculated magnitudes of the cross-section may indicate that error estimates are overly conservative. The slight horizontal shift between the measured and calculated cross-sections could be attributed to inaccuracies of either the $\textit{ab initio}$ molecular potentials or the dipole moment function. Given that the current calculation is non-relativistic this shift is not surprising.

\begin{table}[b]
\caption{Spectroscopic molecular constants for the X$^1\Sigma^+$ and A$^1\Pi$ potentials of both SrCl$^+$ and BaCl$^+$.}
\vspace*{0.2cm}
\begin{tabular}{l|llrrl} \hline
Ion                &  State          &  $R_e$  &$D_e/(hc)$  & $\omega_e/(hc)$ & $B_e/(hc)$   \\
                     &                      & ($a_0$) & (cm$^{-1}$) & (cm$^{-1}$) & (cm$^{-1}$) \\
\hline
 \multirow{2}{*}{SrCl$^+$}   & X$^1\Sigma^+$  & 4.69          & 34158                &  366      &  0.1096     \\
                      & A$^1\Pi$            & 6.32           & 1658                  &  77        &  0.0603     \\
\hline

 \multirow{2}{*}{BaCl$^+$}    &  X$^1\Sigma^+$  & 4.89         & 39103                &   329      &  0.0903    \\
                      & A$^1\Pi$             &  6.38         & 2083                 &   96         & 0.0528      \\
\hline
\end{tabular}
\label{constants}
\end{table}

\begin{figure}
\includegraphics[scale=.9]{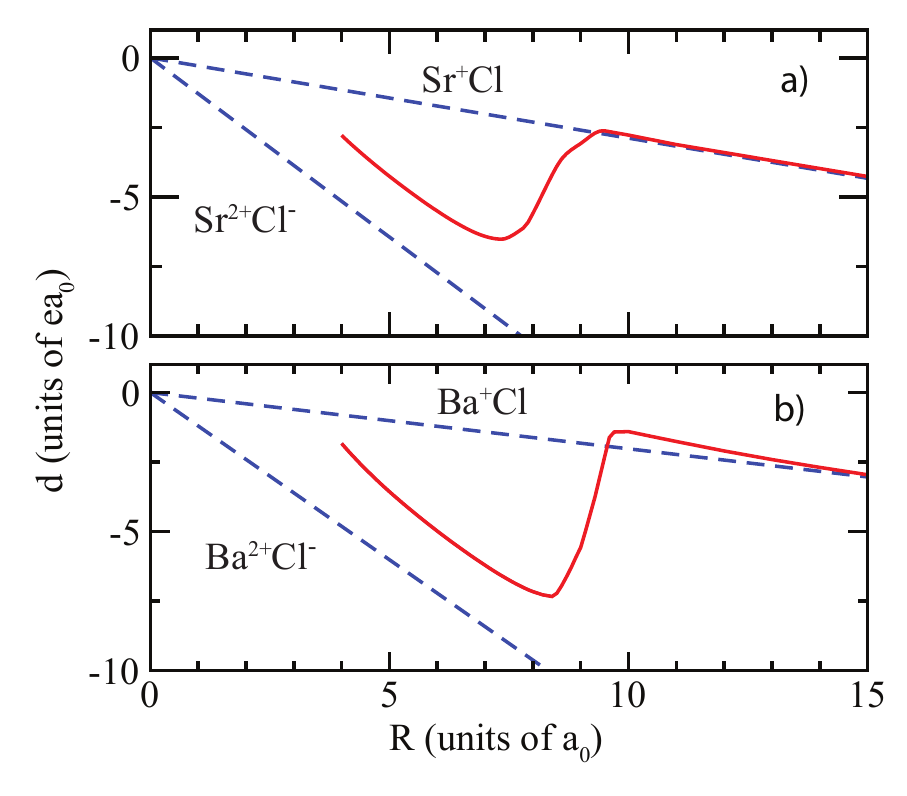}
\caption{The permanent electric dipole moment (solid lines) of the X$^1\Sigma^+$ state as a function of internuclear separation for SrCl$^+$ (panel a) and BaCl$^+$ (panel b).   The dipole moment is given relative to the center of mass. The top dashed line in each panel corresponds to the classical
dipole moment for a single positive charge placed at the position of the barium or strontium ion, respectively.  The bottom dashed line corresponds to the classical
dipole moment for a doubly charged barium or strontium ion and a negatively charged chlorine anion}
\label{pmd}
\end{figure}

The spectroscopic constants for the states relevant to this work are give in Table \ref{constants}; constants~\cite{Rellergert2013a} for the same states in BaCl$^+$ are also presented for comparison. The equillibrium bond lengths for these molecules, and therefore the rotational constants, are very similiar. However, we observe that the ground and first excited state potentials of SrCl$^+$  are less deep than those of BaCl$^+$. To understand this relationship, we analyze the permanent dipole moments and the charge distributions for the ground state of both molecules. Figure \ref{pmd} shows the permanent dipole moment of the X state of BaCl$^+$ and SrCl$^+$ as a function of internuclear separation $R$ defined relative to the center of mass of each molecule, i.e. its center of rotation.  The two dashed lines in either panel correspond to the dipole moment of two limiting charge distributions. The ones labeled by Ba$^+$Cl and Sr$^+$Cl correspond to a dipole moment for a singly charged barium or strontium ion and a neutral chlorine atom. The other dashed line in either panel corresponds to the dipole moment for a doubly charged barium or strontium ion and a negatively charged chlorine anion.  

Our calculations show that for R $\le 8a_0$ for SrCl$^+$ and R $\le 9a_0$ for BaCl$^+$, the electronic wavefunction is ``ionic''
in character, dominated by configurations that contain the closed shell Sr$^{2+}$ and Cl$^{-}$ and Ba$^{2+}$ and Cl$^{-}$ ions, respectively. At separations larger than 8$a_0$  or  9$a_0$, respectively, the electronic wavefunctions rapidly change to a covalently bonded state with neutral chlorine and ionic strontium and barium. The numerical dipole moments tend to the dipole moments predicted by the corresponding limiting charge distributions. Because the double ionization potential~\cite{Kramida2013} of Sr is larger than that of Ba, we expect the ionic coupling of BaCl$^+$ to be stronger than that of SrCl$^+$, which is supported by the relative difference in photodissociation energies of these molecules.

\section{Summary}

We have described the use of an integrated linear ion trap and time-of-flight mass spectrometer to record the photodissociation cross-section of SrCl$^+$ in the spectral range of 36000 -- 46000~cm$^{-1}$. \textit{Ab initio} molecular potentials and transition moments were calculated. These calculations indicate that the A$^1\Pi\leftarrow$ X$^1\Sigma^+$ transition is responsible for the observed photodissociation signal and show good agreement with the measurement. Based on these calculations spectroscopic constants for the lowest two electronic states of SrCl$^+$ were reported.

As outlined in the introduction, a new effort for producing and studying ultracold molecular ions is rapidly emerging in physics and chemistry. This effort, despite notable results, is hampered by a significant lack of small molecular ion spectroscopic data. The work presented here provides the first spectroscopic data for SrCl$^+$, an interesting candidate for these studies. It also outlines the design and use of an apparatus which overcomes many of the challenges of molecular ion spectroscopy and should be applicable to higher resolution spectroscopic studies, e.g  pre- and multi-photon dissociation spectroscopy. Nonetheless, for the field of ultracold molecular ions to reach its full potential a renewed effort from the community in small molecular ion spectroscopy is required. 

\section{Acknowledgments}

The authors thank Shylo Stiteler for his assistance with the machining of the stainless steel TOF-MS device; the instrument could not have been constructed without his hard work and mettle. P.P. thanks the U.S. Army Research Office (ARO) Undergraduate Research Apprenticeship Program (URAP) for their support. This work was supported by National Science Foundation (NSF) and ARO grants (Grant Nos. PHY-0855683, PHY-1308573, W911NF-12-1-0476 and W911NF-10-1-0505).

\section{References}

\bibliography{SrCl}
\bibliographystyle{apsrev4-1}

\end{document}